\documentclass[article,preprint,groupedaddress,11pt]{revtex4}
\usepackage{epsfig}
\usepackage{graphicx}
\usepackage{amssymb}

\begin{document}
\title{A proof of the strong cosmic censorship conjecture}
\author{Shahar Hod}
\affiliation{The Ruppin Academic Center, Emeq Hefer 40250, Israel}
\affiliation{ }
\affiliation{The Hadassah Institute, Jerusalem 91010, Israel}
\date{\today}
\centerline {\it This essay is awarded 3rd Prize in the 2020 Essay
Competition of the Gravity Research Foundation}

\begin{abstract}
\ \ \ The Penrose strong cosmic censorship conjecture asserts that
Cauchy horizons inside dynamically formed black holes are unstable
to remnant matter fields that fall into the black holes. The
physical importance of this conjecture stems from the fact that it
provides a necessary condition for general relativity to be a truly
deterministic theory of gravity. Determining the fate of the Penrose
conjecture in non-asymptotically flat black-hole spacetimes has been
the focus of intense research efforts in recent years. In the
present essay we provide a remarkably compact proof, which is based
on Bekenstein's generalized second law of thermodynamics, for the
validity of the intriguing Penrose conjecture in physically
realistic (dynamically formed) curved black-hole spacetimes.
\newline
\newline
Email: shaharhod@gmail.com
\end{abstract}
\bigskip
\maketitle

The physically influential and mathematically elegant singularity theorems of Hawking and Penrose
\cite{HawPen,Pen1,Pen2} have revealed the intriguing fact that the interior regions of
dynamically formed black holes contain spacetime singularities.
These are physically pathological regions in which the Einstein field equations
lose their predictive power.

In order to guarantee the deterministic nature of general relativity
as a successful theory of gravity, Penrose has put forward the
cosmic censorship conjecture, according to which a mysterious
``cosmic censor'' prevents distant observers from being influenced
by the disturbing singularities of highly curved spacetimes
\cite{Pen1,Pen2}. This intriguing conjecture asserts, in particular,
that Cauchy horizons inside physically realistic (dynamically
formed) black holes, which mark the boundaries beyond which the
Einstein field equations lose their predictive power, are singular.
If true, this strong version of the Penrose conjecture would imply
that physical observers are always restricted to live in spacetime
regions in which general relativity is a physically successful and
mathematically deterministic theory of gravity \cite{Pen1,Pen2}.

In order to challenge the validity of the Penrose strong cosmic
censorship conjecture, one may try to identify a physically
realistic curved black-hole spacetime whose inner Cauchy horizon is
stable and regular enough to allow a non-unique continuation of the
inner spacetime into the pathological (non-deterministic) region.
Such pathological black-hole spacetimes, {\it if} exist, would
provide a disturbing counter-example to the cosmic censorship
conjecture and would signal the breakdown of determinism in general
relativity.

The question of the final fate of the Penrose cosmic censorship
conjecture in non-asymptotically flat spacetimes has attracted much
attention from physicists and mathematicians during the last three
years \cite{Refin,Cham,Hod1,Hod1c,Ge}. In particular, it has been
proved \cite{Refin,Cham,Hod1,Hod1c,Ge} that the fundamental nature
(singular/non-singular) of the inner Cauchy horizons in
asymptotically de Sitter black-hole spacetimes is determined by a
delicate interplay between two competing physical mechanisms:
\newline
(1) The characteristic asymptotic {\it decay}
$\psi^{\text{external}}(t\to\infty)\sim e^{-\Im\omega_0\cdot t}$ of
remnant fields in the exterior regions of the dynamically formed black-hole spacetime.
Here $\omega_0$ is the fundamental (least damped) quasinormal resonant frequency which determines
the characteristic relaxation rate of the external spacetime.
\newline
(2) The blue-shift {\it amplification} phenomenon
$\psi^{\text{internal}}(v\to\infty)\sim e^{\kappa_- v}$
\cite{Notevv} which characterizes the dynamics of the infalling
fields as they accumulate along the inner Cauchy horizon of the
dynamically formed black hole. Here $\kappa_-$ is the surface
gravity of the inner black-hole horizon.

Intriguingly, the final fate of the inner Cauchy horizons inside
physically realistic (dynamically formed) black holes in
non-asymptotically flat spacetimes is determined by the simple
dimensionless ratio \cite{Refin}
\begin{equation}\label{Eq1}
\Gamma\equiv {{\Im\omega_0}\over{\kappa_-}}\  .
\end{equation}
In particular, a dynamically formed black hole which is characterized by the dimensionless
inequality $\Gamma>1/2$ contains an inner regular Cauchy
horizon that allows the corresponding spacetime to be continued in non-unique ways \cite{Refin},
thus violating the fundamental Penrose strong cosmic censorship conjecture.

One therefore concludes that the simple inequality
\begin{equation}\label{Eq2}
\Gamma\leq{1\over2}\
\end{equation}
provides a necessary condition for the Einstein field equations
to preserve their predictive power in non-asymptotically flat de Sitter spacetimes.
It is therefore physically important to prove that
the quasinormal resonant spectra of {\it all} dynamically formed
black holes are characterized by the property [see
Eqs. (\ref{Eq1}) and (\ref{Eq2})]
\begin{equation}\label{Eq3}
\Im\omega_0\leq{1\over2}\kappa_-\  .
\end{equation}

It is interesting to note that, using analytical techniques, it has
been proved in \cite{Hod2} that spinning Kerr-de Sitter black-hole
spacetimes respect the inequality (\ref{Eq3}) and therefore respect
determinism. However, the situation is more involved in the case of
charged de Sitter black holes \cite{Refin,Cham,Hod1,Hod1c,Ge,Hodnn}.
In particular, using analytical techniques, it has been explicitly
proved in \cite{Hod1} that composed
charged-de-Sitter-black-holes-charged-massive-fields systems in the
dimensionless physical regime $\mu r_+ \ll qQ \ll (\mu r_+)^2$ [here
$\{\mu,q\}$ are respectively the proper mass and charge coupling
constant of the charged matter field, and $\{Q,r_+\}$ are
respectively the black-hole electric charge and the radius of its
event horizon] are characterized by quasinormal resonant spectra
that respect the fundamental inequality (\ref{Eq3}). One therefore
concludes that these dynamically formed charged black-hole
spacetimes respect determinism (and, in particular, respect the
Penrose strong cosmic censorship conjecture).

For most physicists, the explicit calculations presented in
\cite{Hod1} for the validity of the necessary inequality (\ref{Eq3})
\cite{Notenec} in the physical regime $\mu r_+ \ll qQ \ll (\mu
r_+)^2$ are certainly not enough. In particular, in order to prove
the validity of the Penrose conjecture for all black-hole
spacetimes, it is necessary to have a generic (that is,
parameter-independent) proof that {\it all} physically realistic
(dynamically formed) black holes are characterized by quasinormal
relaxation spectra that respect the fundamental inequality
(\ref{Eq3}).

In view of the complex mathematical nature of the
Einstein-charged-matter field equations
\cite{Refin,Cham,Hod1,Hod1c,Ge}, it may seem that any direct attempt
to obtain a general proof for the validity of the strong cosmic
censorship conjecture in charged black-hole spacetimes is doomed to
fail. In particular, it should be realized that a direct test of the
Penrose conjecture in charged de Sitter black-hole spacetimes would
require one to scan numerically the infinitely large phase space of
the black-hole-field physical parameters
$\{r_-,r_+,r_{\text{c}},q,\mu\}$ \cite{Noteradii} in search of a
dynamically formed charged black hole that violates the necessary
inequality (\ref{Eq3}) and therefore violates the strong cosmic
censorship conjecture \cite{Notenec}. It is clear that this direct
numerical approach to the problem is a truly time consuming
Sisyphean task!

But we need not lose heart. Our experience in physics has taught us
that the fundamental laws of nature may sometimes provide, in
remarkably elegant ways, important insights about the physical
properties of highly complex systems. Bekenstein's generalized
second law of thermodynamics is among these truly remarkable laws
\cite{Bek1}. It states that the total entropy of a black-hole
spacetime is a non-decreasing quantity in self-consistent quantum
theories of gravity.

Intriguingly, and most importantly from the point of view of the
strong cosmic censorship conjecture, it has been explicitly proved
in \cite{Hod3} that the Bekenstein generalized second law of
thermodynamics \cite{Bek1} implies that the characteristic
relaxation time $\tau$ of a perturbed thermodynamic system is
bounded from below by the remarkably compact time-times-temperature
(TTT) quantum relation \cite{Hod3}
\begin{equation}\label{Eq4}
\tau\times T\geq {{\hbar}\over{\pi}}\  ,
\end{equation}
where $T$ is the characteristic temperature of the physical system.

Black-hole spacetimes, like mundane thermodynamic systems, are known
to be characterized by a well defined temperature, which is given by
the famous Bekenstein-Hawking relation \cite{Bek1,Haw1}
\begin{equation}\label{Eq5}
T_{\text{BH}}={{\kappa_+}\over{2\pi}}\cdot\hbar\  ,
\end{equation}
where $\kappa_+$ is the surface gravity of the black-hole event
horizon. Substituting the Bekenstein-Hawking black-hole temperature
(\ref{Eq5}) into the universal relaxation bound (\ref{Eq4}) and
using the relation $\tau_{\text{relax}}\equiv 1/\Im\omega_0$ for the
characteristic relaxation time of the dynamically formed black-hole
spacetime, one finds that the quasinormal resonant spectra of {\it
all} physically realistic (dynamically formed) black-hole spacetimes
are characterized by the compact upper bound
\begin{equation}\label{Eq6}
\Im\omega_0\leq {1\over2} \kappa_+\  .
\end{equation}
We have therefore established a remarkably simple (and physically
important) relation between the characteristic relaxation rates of
dynamically formed black-hole spacetimes and the corresponding
black-hole surface gravities.

{\it Summary.---} The Penrose strong cosmic censorship conjecture
\cite{HawPen,Pen1,Pen2}, which asserts that general relativity is a
deterministic theory of gravity, has attracted much attention from
physicists and mathematicians during the last five decades. In
particular, the final fate of this conjecture in non-asymptotically
flat charged black-hole spacetimes has been the focus of an intense
debate during the last three years \cite{Refin,Cham,Hod1,Hod1c,Ge}.

The Penrose conjecture states that the inner spacetime regions of
physically realistic (dynamically formed) black holes cannot be
extended in a non-deterministic (non-unique) way beyond their inner
Cauchy horizons. The conjecture therefore implies that the inner
horizons of black-hole spacetimes must be unstable to remnant matter
fields that fall into the dynamically formed black holes. In
particular, the relaxation-rate-inner-surface-gravity relation
$\Im\omega_0\leq{1\over2}\kappa_-$ [see Eq. (\ref{Eq3})] provides a
necessary condition for the validity of the Penrose conjecture in
asymptotically de Sitter black-hole spacetimes
\cite{Refin,Cham,Hod1,Hod1c,Ge}.

We have emphasized the fact that a direct (brute force) approach to
test the validity of the Penrose strong cosmic censorship conjecture
[and the closely related fundamental inequality (\ref{Eq3})] in
physically realistic black-hole spacetimes would be to scan
numerically the infinitely large phase space of the quasinormal
resonant spectra which characterize the late-time relaxation of the
dynamically formed black holes.

Instead of following this truly Sisyphean approach, we have pointed
out that the Bekenstein generalized second law of thermodynamics
\cite{Bek1} implies that thermodynamic systems with well defined
temperatures, including black holes, are characterized by the
universal relaxation bound $\Im\omega_0\leq {1\over2} \kappa_+$ [see
Eq. (\ref{Eq6})] \cite{Hod3}. Using the characteristic inequality
$\kappa_+\leq \kappa_-$ for the black-hole surface gravities
\cite{Cham}, one concludes that the relaxation spectra of {\it all}
dynamically formed black-hole spacetimes are characterized by the
relaxation-rate-surface-gravity inequality
$\Im\omega_0\leq{1\over2}\kappa_-$. This fact implies that the
corresponding inner Cauchy horizons are dynamically unstable to
infalling remnant fields \cite{Refin,Cham,Hod1,Hod1c,Ge}. Thus, the
inner black-hole spacetime cannot be extended in a non-unique
(non-deterministic) way beyond these horizons.

We have therefore proved that physically realistic (dynamically
formed) black-hole spacetimes {\it respect} the Penrose strong
cosmic censorship conjecture.

\newpage

\noindent
{\bf ACKNOWLEDGMENTS}
\bigskip

This research is supported by the Carmel Science Foundation. I would
like to thank Yael Oren, Arbel M. Ongo, Ayelet B. Lata, and Alona B.
Tea for helpful discussions.



\begin{thebibliography}{99}

\bibitem{HawPen} S. W. Hawking and R. Penrose, Proc. R. Soc. Lond. {\bf A314}, 529 (1970).

\bibitem{Pen1} R. Penrose, Riv. Nuovo Cimento I {\bf 1}, 252 (1969).

\bibitem{Pen2} R. Penrose in {\it General Relativity, an Einstein Centenary
Survey}, eds. S.W. Hawking and W. Israel (Cambridge University
Press, 1979).

\bibitem{Refin} See details in \cite{Cham,Hod1,Hod1c,Ge} and references therein.

\bibitem{Cham} C. Chambers, arXiv:gr-qc/9709025.

\bibitem{Hod1} S. Hod, Nucl. Phys. B {\bf 941}, 636 (2019) [arXiv:1801.07261].

\bibitem{Hod1c} S. Hod, arXiv:1810.04853.

\bibitem{Ge} B. Ge, J. Jiang, B. Wang, H. Zhang, and Z. Zhong, JHEP {\bf 01}, 123 (2019) [arXiv:1810.12128].

\bibitem{Notevv} We use gravitational units in which $G=c=1$. Here $v$ is the standard advanced null coordinate.

\bibitem{Hod2} See S. Hod, Phys. Lett. B {\bf 780}, 221 (2018) [arXiv:1803.05443] and references therein.

\bibitem{Hodnn} S. Hod, Nucl. Phys. B {\bf 948}, 114772 (2019) [arXiv:1910.09564].

\bibitem{Notenec} It is worth emphasizing again that the inequality (\ref{Eq3}) serves
as a necessary condition for the validity of the Penrose strong
cosmic censorship conjecture in non-asymptotically flat black-hole
spacetimes \cite{Refin}.

\bibitem{Noteradii} Here $\{r_-,r_+,r_{\text{c}}\}$ are respectively
the inner (Cauchy), outer (event), and cosmologival horizons of the
non-asymptotically flat charged black-hole spacetime.

\bibitem{Bek1} J. D. Bekenstein, Phys. Rev. D {\bf 7}, 2333 (1973).

\bibitem{Hod3} S. Hod, Phys. Rev. D {\bf 75}, 064013 (2007) [arXiv:gr-qc/0611004].

\bibitem{Haw1} S. W. Hawking, Commun. Math. Phys. {\bf 43}, 199 (1975).

\end{thebibliography}
\end{document}